\documentclass[linenumbers]{elsarticle}
\usepackage{graphicx}
\usepackage{amsmath}
\usepackage{units}
\usepackage{amssymb}
\usepackage{subfigure}
\usepackage{fixltx2e} 
\usepackage{verbatim} 
\usepackage[numbers]{natbib} 

\usepackage{lineno}
\linenumbers

\begin{document} 

\title{Design of a wavelength frame multiplication system using acceptance diagrams}
\date{\today}

\author[hzb,bmbf]{D. Nekrassov\corref{cor1}} 
\ead{daniil.nekrassov@helmholtz-berlin.de}
\author[hzb,bmbf]{C. Zendler} 
\ead{carolin.zendler@helmholtz-berlin.de}
\author[hzb,bmbf]{K. Lieutenant}
\ead{klaus.lieutenant@helmholtz-berlin.de}
\address[hzb]{Helmholtz-Zentrum Berlin, Hahn-Meitner-Platz 1, D-14109 Berlin, Germany}
\address[bmbf]{German Work Package for the ESS Design Update}
\cortext[cor1]{Corresponding author}

\begin{abstract}

The concept of Wavelength Frame Multiplication (WFM) was developed to extend the usable wavelength range on long pulse neutron sources for instruments using pulse shaping choppers. For some instruments, it is combined with a pulse shaping double chopper, which defines a constant wavelength resolution, and a set of frame overlap choppers that prevent spurious neutrons from reaching the detector thus avoiding systematic errors in the calculation of wavelength from time of flight. Due to its complexity, the design of such a system is challenging and there are several criteria that need to be accounted for. In this work, the design of the WFM chopper system for the potential future liquids reflectometer at the European Spallation Source (ESS) is presented, which makes use of acceptance diagrams. They prove to be a powerful tool for understanding the work principle of the system and recognizing potential problems. The authors assume that the presented study can be useful for design or upgrade of further instruments, in particular the ones planned for the ESS.

\end{abstract}

\maketitle

\section{Introduction}

There is currently an increasing demand for neutron instruments, at which the resolution can be adjusted, in particular towards high-resolution setups. The total instrument resolution in neutron scattering experiments always depends, amongst others, on the experimental $\delta \lambda / \lambda$ resolution, where $\lambda$ is the neutron wavelength. In time-of-flight (ToF) mode, the experimental resolution is determined by pulse shaping choppers for all instruments at continuous sources and for high or medium resolution on long pulse sources. A particular system of rotating disc choppers provides the desired waveband and removes contaminant neutrons. For some experiments like small-angle neutron scattering or neutron reflectometry, it is often desirable to have a constant wavelength resolution over the entire usable waveband. For reactor sources, this can be achieved by introducing a pulse shaping double chopper operating in optically blind mode \cite{Bib:vanWell}. In this case, the wavelength resolution is determined by the ratio of the distance $D$ between the pulse shaping choppers and the distance $L_0$ between the center of the double chopper system and the detector: $\delta \lambda / \lambda = D / L_0$. This relation is valid for all wavelengths up to $\lambda = \frac{3956}{D/\tau} [\mathrm{\AA}]$, where $\tau$ is the single disc opening time.\\

At pulsed sources, like the currently planned European Spallation Source (ESS) \cite{Bib:ESS}, the chopper design described above \cite{Bib:vanWell} is usually not applicable in its simple form. The reason is that due to the needed shielding volume, the first chopper can be placed only at a certain minimum distance away from the source, which is currently 6\,m for the ESS. Depending on the desired waveband, this implicates that not all neutrons will be at the first chopper at the same time, which limits the usable waveband at the detector. To extend this range, the WFM concept was developed \cite{Bib:WFM}. It was then complemented with a blind double-chopper setup to create a wavelength dependent pulse length \cite{Bib:WFM2}. Here, the combination with a blind double-chopper setup is used to obtain a constant wavelength resolution. To achieve a sufficiently broadband pulse within the main frame (given by the pulse repetition rate), this concept utilizes multiple subframes. These subframes are constructed such that the wavelength resolution is the same for every subframe and they are separated in time at the detector, but at the same time the measurement time is efficiently used, i.e. the time gaps between individual subframes are minimised. The proof of principle of the WFM approach was achieved at the Budapest Neutron Center (BNC) \cite{Bib:WFMExp}. \\

At the future ESS, several instruments will need to implement the WFM approach. The chopper layout must be carefully adapted to the long pulse structure of the ESS beam. Neutrons being detected in the wrong subframe can pose a significant source of systematic errors \footnote{or spoil some fraction of the dataset and thereby lengthen the measurement time, if a contaminated part of a subframe has to be removed from the later data analysis.}, so in particular the choice of frame overlap chopper parameters must be done with great care. The need for a thorough analysis method was lastly shown by several technical challenges experienced during the conception of a WFM chopper layout using time-of-flight diagrams for the ESS test beamline in Berlin \cite{Bib:WFMTestBeamLine}. In this paper, the design of a WFM setup carried out in the context of a design study of a liquids reflectometer to be proposed for the ESS, is demonstrated by using acceptance diagrams based on the work presented in \cite{Bib:AccDiag}.

\section{Application of acceptance diagrams for WFM system of the ESS liquids reflectometer}

\subsection{Designing the pulse shaping choppers}

\begin{figure}
\begin{center}
	\includegraphics[width=0.95\textwidth]{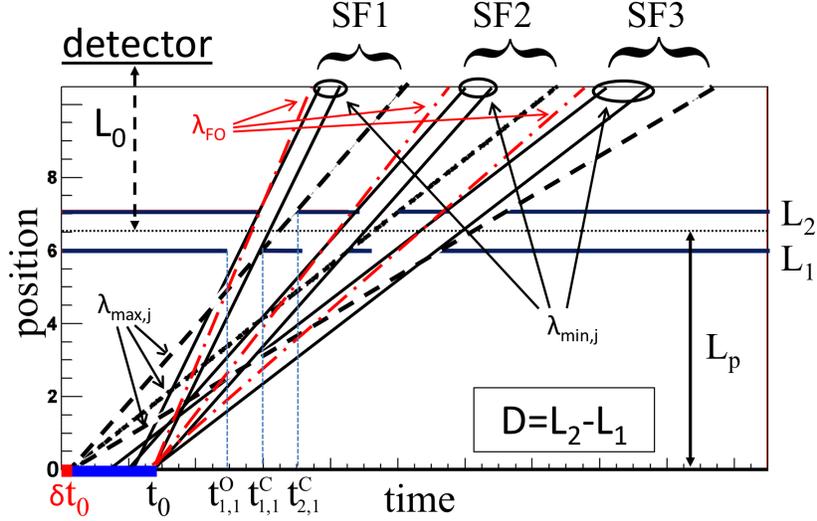}
	\caption{Illustration of the construction procedure of the PSC with a ToF diagram. The total pulse duration $t_0$ is denoted by the blue bar, while the time offset $\delta t_0$ is illustrated by the red square, thus the usable pulse length is $t_0 - \delta t_0$. The choppers are located at the positions L$_1$ and L$_2$. For the $j$th subframe SF, neutrons having the wavelength $\lambda_{{\mathrm{min}}, j}$ and $\lambda_{{\mathrm{max}}, j}$ used in Eqs. \ref{Eq:FirstWindowCloseCh1} and \ref{Eq:FirstWindowCloseCh2} are shown by black lines. Neutrons with wavelengths $\lambda_{\mathrm{FO}} < \lambda_{\mathrm{min}, j}$ responsible for potential subframe overlap, are depicted by dashed-dotted red lines. In addition, the chopper system parameters $D$ being the distance between both PSCs, the distance between the source and the centre of the PSC system $L_\mathrm{p}$ and $L_0$, which is the distance between the centre of the PSC system and the detector that is well outside the illustrated region, are also shown. See text for further details.}
\label{Fig:PSCDiagram}
\end{center}
\end{figure}

In a WFM chopper setup, the parameters of the pulse shaping choppers (PSCs) have to be calculated first. These depend on the global parameters being the total length $L_{\mathrm{tot}}$ of the instrument and the width of the waveband $\Delta \lambda = \lambda_{\mathrm{max}} - \lambda_{\mathrm{min}}$, where $\lambda_{\mathrm{min}}$ and $\lambda_{\mathrm{max}}$ are the minimal and maximal design wavelengths, respectively. The instrument length and the waveband width are related through the source period T:

\begin{equation}
\Delta \lambda = h / m_n \times T / L_{\mathrm{tot}},
\label{Eq:WidthWaveband}
\end{equation}

where $h$ is Planck's constant and $m_n$ is the neutron mass. In addition, it is important to decide on the loosest wavelength resolution $R_{\mathrm{max}}=(\delta \lambda / \lambda)_{\mathrm{max}}$ in the WFM regime. Once these parameters are given, then the distance $D = L_0 \times (\delta \lambda / \lambda)_\mathrm{max}$ between the two choppers, the number of windows, their sizes and offsets with respect to each other can be calculated (see Fig. \ref{Fig:PSCDiagram}). The windows of the PSCs are designed such that they enable measurements with the loosest design resolution $R_{\mathrm{max}}$, with the distance between the two choppers being

\begin{equation}
D = L_2 - L_1 = L_0 \times R_{\mathrm{max}},
\label{Eq:D_Def}
\end{equation}

where $L_1$ ($L_2$) is the position of the first (second) PSC chopper. Higher resolutions are then achieved by reducing the distance between the two choppers \cite{Bib:vanWell}.\\

The design of the chopper windows starts by calculating the time $t^{\mathrm{C}}_{1,1}$ when the first window ($W_{1,1}$) of the first chopper Ch$_1$ closes. This time is set by neutrons of wavelength $\lambda_{\mathrm{min}}$ starting at the end of the pulse, see Fig. \ref{Fig:PSCDiagram}:

\begin{equation}
t^{\mathrm {C}}_{1,1} = L_1 / v(\lambda_{\mathrm{min}}) +  t_0
\label{Eq:FirstWindowCloseCh1}
\end{equation}

The PSCs operate in the optical blind mode, i.e. the second chopper opens when the first one closes. Thus $t^{\mathrm {O}}_{2,1} = t^{\mathrm {C}}_{1,1}$. The opening time $t^{O}_{1,1}$ of the window $W_{1,1}$ is then given by the slowest neutrons that can reach the second chopper when the window $W_{2,1}$ opens, which start at the source at the beginning of the pulse or after some offset $\delta t_0$:

\begin{equation}
t^{O}_{1,1} = \frac{L_1}{\check{v}_1} + \delta t_0, 
\label{Eq:FirstWindowOpenCh1}
\end{equation}

where $\check{v}_1 = L_{2}/(t^{\mathrm {O}}_{2,1} - \delta t_0)$. The closing time $t^{C}_{2,1}$ of the window $W_{2,1}$ is given by the slowest neutrons with the wavelength $\lambda_{\mathrm{max}, 1}$ that reach the first chopper when it closes: 

\begin{equation}
t^{C}_{2,1} = \frac{L_2}{v_{\mathrm {min}, 1}} + \delta t_0, 
\label{Eq:FirstWindowCloseCh2}
\end{equation}

where $v_{{\mathrm {min}}, 1} = L_{1}/(t^{\mathrm {C}}_{1,1} - \delta t_0)$. Note that $\lambda_{\mathrm{min}}$ is not the shortest wavelength that gets transmitted through the PSC (see Fig. \ref{Fig:PSCDiagram}), but is the shortest wavelength for which the created pulse length $\delta t$ corresponds to the resolution $R_{\mathrm{max}}$. At the same time, if $\delta t_0 > 0$, then $\lambda_{\mathrm{max}, 1}=\lambda(v_{{\mathrm {min}}, 1})$ is also not the largest wavelength that gets transmitted through the first window of the PSCs. For the design of the second window, the shortest wavelength is set $\lambda_{\mathrm{min}, 2} = \lambda_{\mathrm{max}, 1}$ to achieve a continuous spectrum and minimise time gaps at the detector, and the construction procedure is repeated iteratively. Thus neutrons with wavelengths $\lambda < \lambda_{\mathrm{min}, j}$ or $\lambda > \lambda_{\mathrm{max}, j}$ that get transmitted through the $j$th window of the PSCs can lead to overlap of the subframes in time at some distance behind the PSCs and must be treated by frame overlap choppers. Their design is discussed in the next subsection. \\

\begin{figure}
\begin{center}
        \subfigure[Total phase space at the PSC]{\includegraphics[width=0.485\textwidth,keepaspectratio]{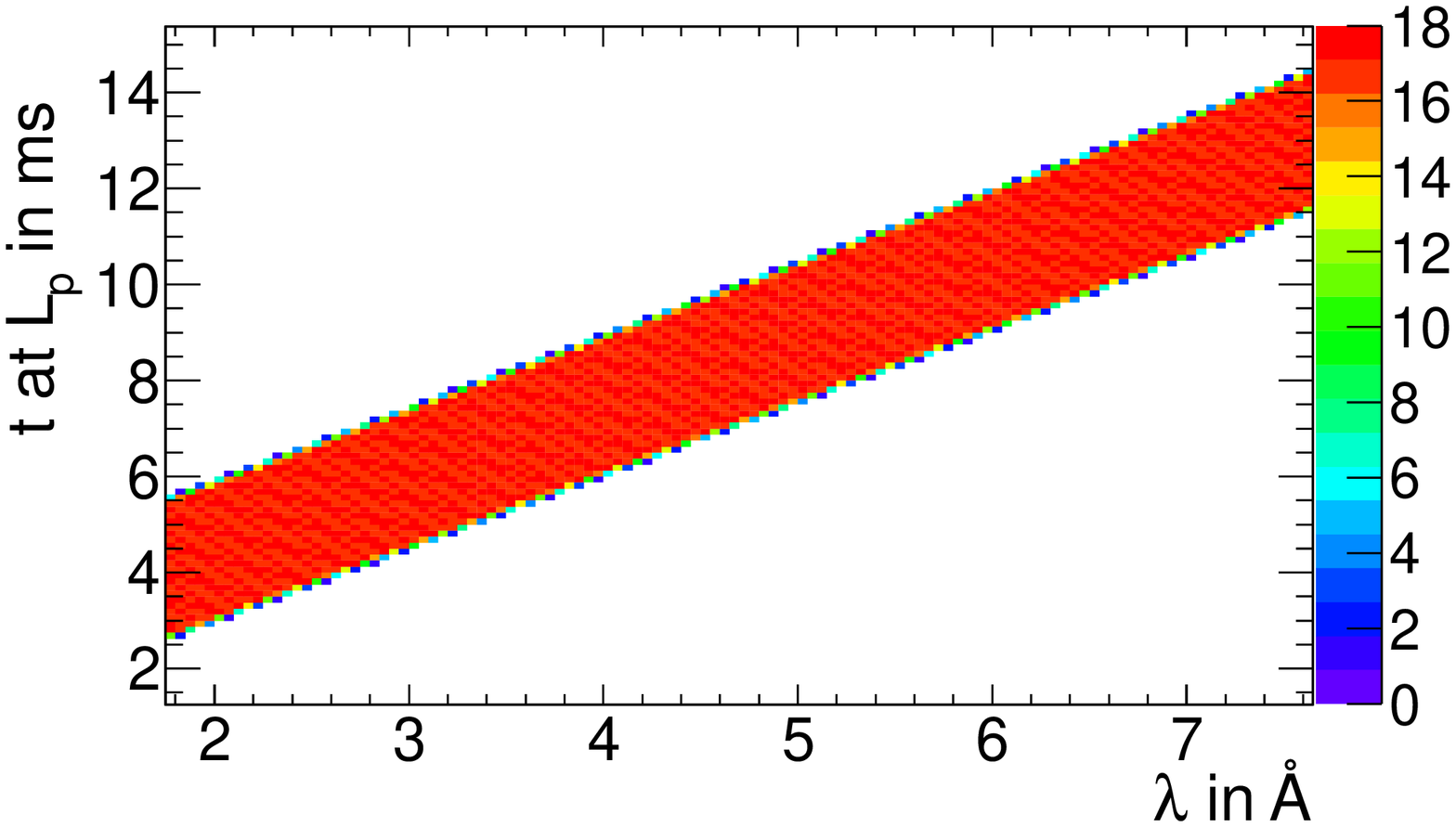}}
        \subfigure[Phase space after shaping by the PSC]{\includegraphics[width=0.485\textwidth,keepaspectratio]{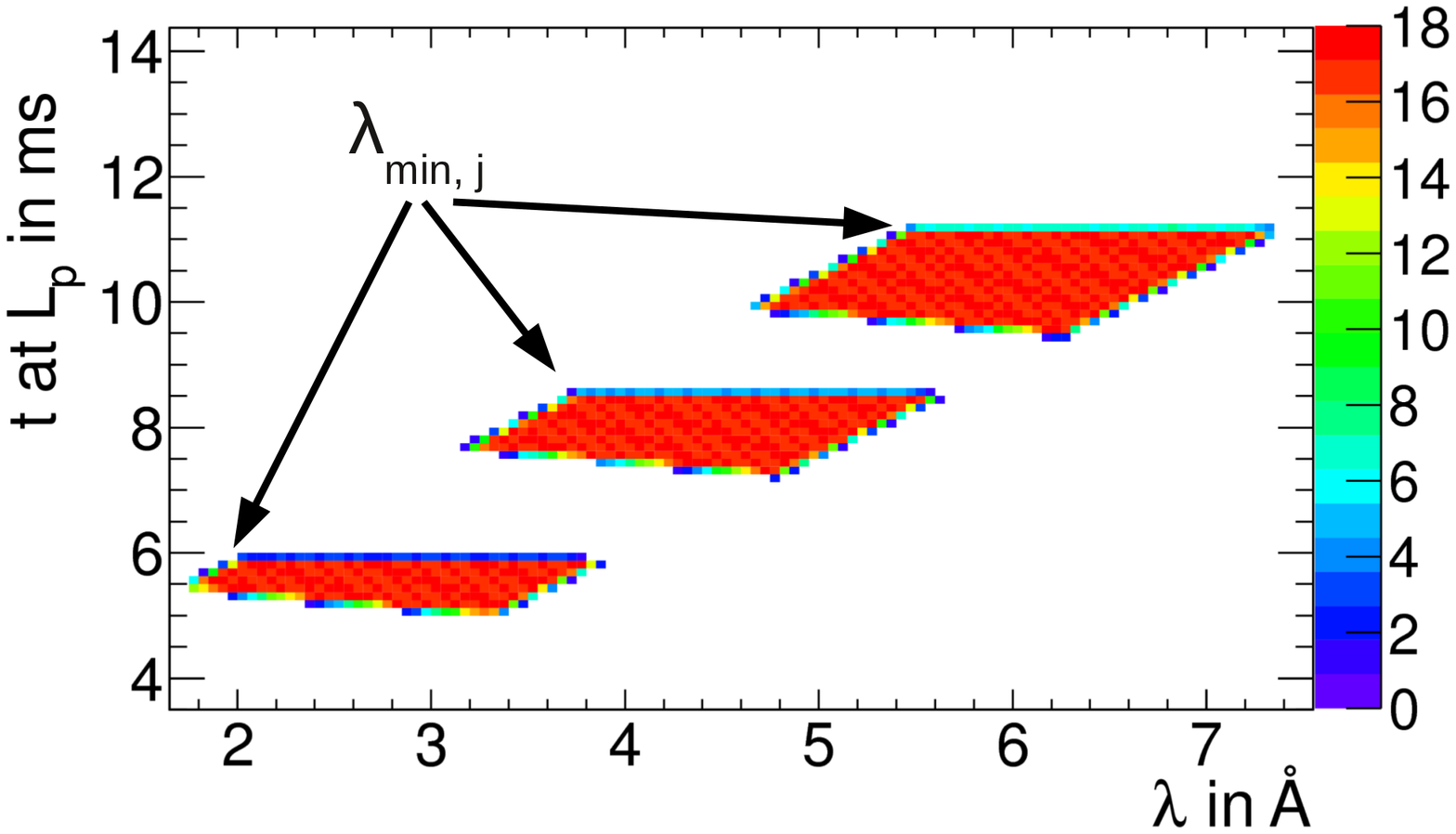}}
        \caption{Neutron phase space available at the PSC for the \textit{instrument I}, displayed as correlation between the neutron wavelength $\lambda$ and the ToF at the position between the PSCs. This phase space has been determined by a fixed grid scan through the $[\lambda, t]$ parameter space. The units on the z-axis are arbitrary and correspond to the phase space density. Without any pulse shaping, the phase space is linearly correlated and has the ESS pulse width of 2.86\,ms for each wavelength. After pulse shaping, the phase space is divided into three subframes, with the width $\delta t (\lambda)$ corresponding to the design resolution.}
\label{Fig:PhaseSpaceNoFOC}
 \end{center}
\end{figure}

A PSC constructed in the way described above transmits a certain fraction of the total available phase space. The latter is obtained by performing a fixed grid scan through the $[t, \lambda]$ parameter space assuming a constant spectrum as a function of the wavelength $\lambda$, where $t$ is the start time of a neutron at the source. This can be visualised in an acceptance diagram (Fig. \ref{Fig:PhaseSpaceNoFOC}) displaying the correlation between the neutron wavelength $\lambda$ and the time $t_{\mathrm{PS}}$, at which the neutron is at the position $L_\mathrm{p} = L_1 + 1/2 \times D$ located in the center between both PSCs. As an example, instrument parameters calculated for a potential ESS liquids reflectometer (\textit{instrument I}) (see Table \ref{Tab:LR}) are used in the most of the following discussion. The initially available phase space is split by the PSCs into 3 subframes being disjoint in time but joint in wavelength ranging from $2 \, \mathrm{\AA}$ to $7.2\, \mathrm{\AA}$, based on a instrument length of $L_{\mathrm{tot}} = 55 \, \mathrm{m}$. For each $\lambda$, the total width $\delta t (\lambda)$ of the modified pulse corresponds to the design resolution $2.2\%$ of the WFM system. If no further choppers would be included in the system, due to wavelength overlap of individual subframes discussed above, the subframes would inevitably overlap in time at some distance after the PSC. Thus frame overlap choppers are needed to keep the subpulses separated until they reach the detector. Their number and positions are optimised in the following using acceptance diagrams. \\

\subsection{Designing the frame overlap choppers}

Frame overlap choppers (FOCs) can be visualised in the acceptance diagram as linear functions indicating the opening and closing of the corresponding chopper window. Points in the phase space described by these functions correspond to certain $[t, \lambda]$ combinations such that these neutrons reach the corresponding chopper at the time when it opens or closes. The analytical description of these functions for the opening and closing time is: 

\begin{equation}
\begin{split}
t_{i,j}^{O/C} = f(\lambda) & = -((L_i - L_\mathrm{p}) / v(\lambda))  + \Theta^{O/C}_{i,j} / \omega_i \\
                           & = -((L_i - L_\mathrm{p})\times m/h)\times \lambda  + \Theta^{O/C}_{i,j} / \omega_i,
\end{split}
\label{Eq:FOCAccDiag}
\end{equation}

where $L_i$ is the distance between the Chopper $i$ and the source, $v(\lambda) = \frac{h}{m \lambda}$ the neutron velocity, $\Theta^{O/C}_{i, j}$ the angular offset of the window start (end) $j$ with respect to the guide position and $\omega_i$ the chopper rotation frequency. At a pulsed source, chopper frequencies have to be equal to the source frequency or larger by an integer factor. Fractional distances between the PSCs\footnote{or the source if the pulse is not shaped afterwards.} and the detector act thereby as a limit for maximum possible multiple of the source frequency, e.g. choppers only can rotate at twice (four times) the source frequency, if their distance $D_i$ to the PSCs fulfills $D_i \leq 1/2 L_0 \, (1/4 L_0)$ and so on. Thus as a first choice, three FOCs can be placed at $1/8 L_0 +  L_1 = 12.125 \, \mathrm{m}$, $1/4 L_0 +  L_1 = 18.25 \, \mathrm{m}$ and $1/2 L_0 +  L_{1} = 30.5 \, \mathrm{m}$. The windows of a FOC $i$ are then constructed such that they open when they are reached by the fastest neutron starting at $t^{\lambda_{\mathrm{min},j}}_{j} = t^{O}_{2,j} - L_2/v(\lambda_{\mathrm{min}, j})$ and close upon arrival of the slowest neutron of the corresponding subframe $j$ starting at $\delta t_0$. Based on these foregoing considerations, the window parameters $j$ of the FOC $i$ can be calculated in a straightforward way: 

\begin{equation}
\Theta^{O}_{i, j} = -\omega_i \times (\frac{L_i}{v(\lambda_{\mathrm{min}, j})} + t^{\lambda_{\mathrm{min}, j}}_j)
\label{Eq:FOCStart}
\end{equation}

\begin{equation}
\Theta^{C}_{i, j} = -\omega_i \times (\frac{L_i}{v(\lambda_{\mathrm{max}, j})} + \delta t_0)
\label{Eq:FOCEnd}
\end{equation}

\begin{figure}
\begin{center}
        \subfigure[Phase space after inclusion of FOC1 at $12.125\,$m]{\includegraphics[width=0.485\textwidth,keepaspectratio]{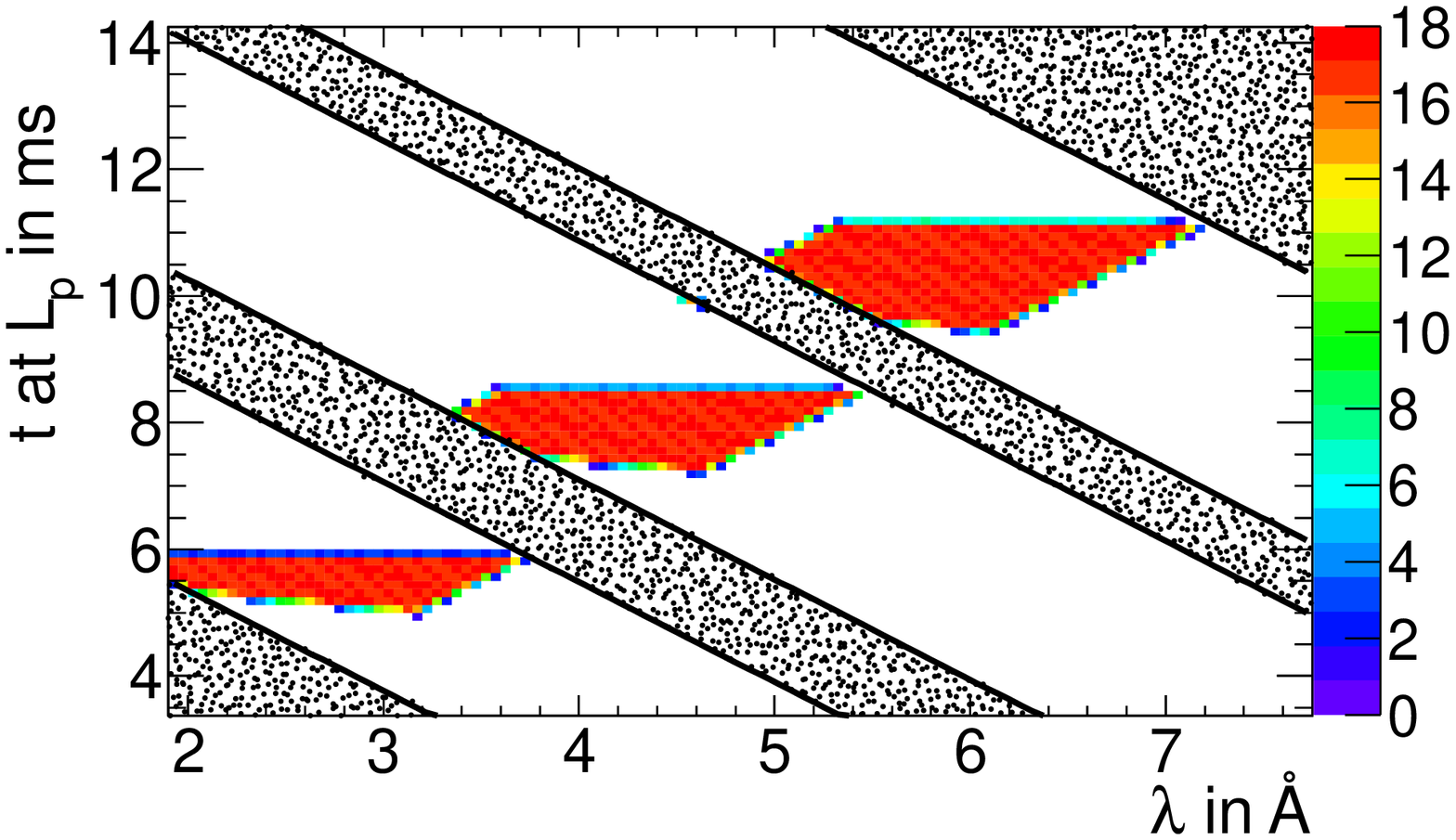}}
        \subfigure[Phase space after inclusion of FOC2 at $18.25\,$m]{\includegraphics[width=0.485\textwidth,keepaspectratio]{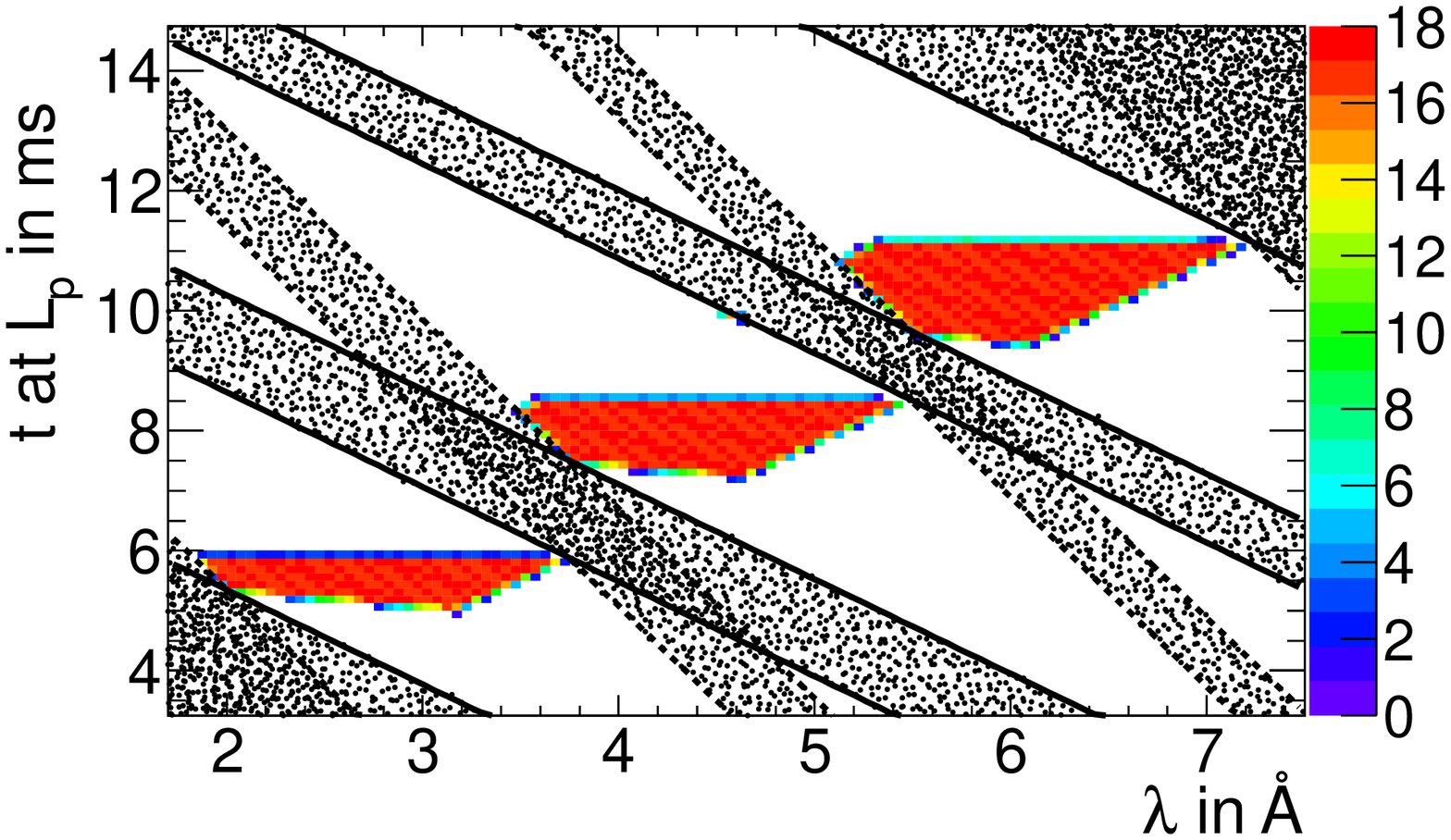}}
	\hfill
	\subfigure[Phase space after inclusion of FOC3 at $30.5\,$m]{\includegraphics[width=0.485\textwidth,keepaspectratio]{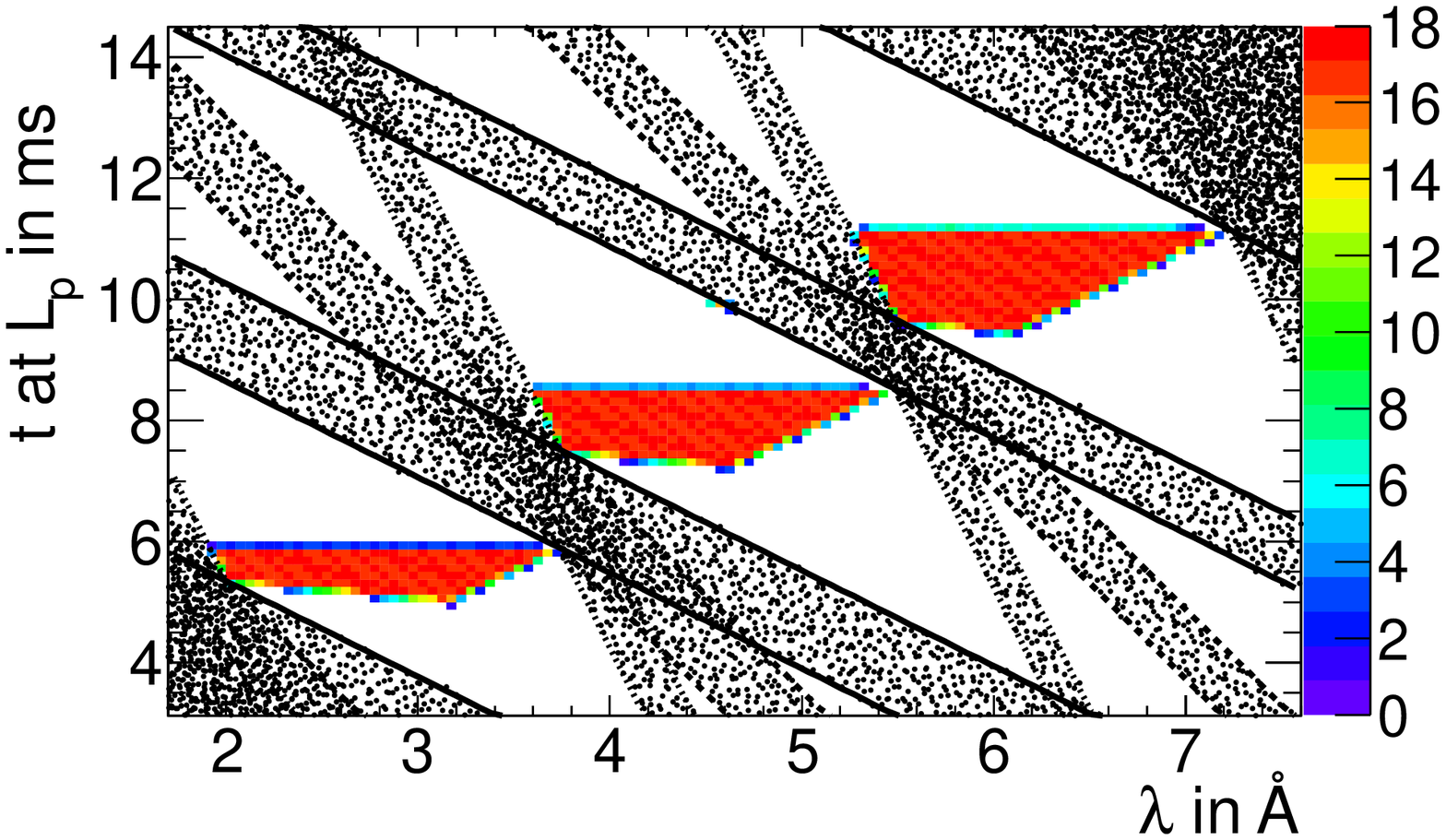}}
        \subfigure[Phase space for $1\%$ resolution]{\includegraphics[width=0.485\textwidth,keepaspectratio]{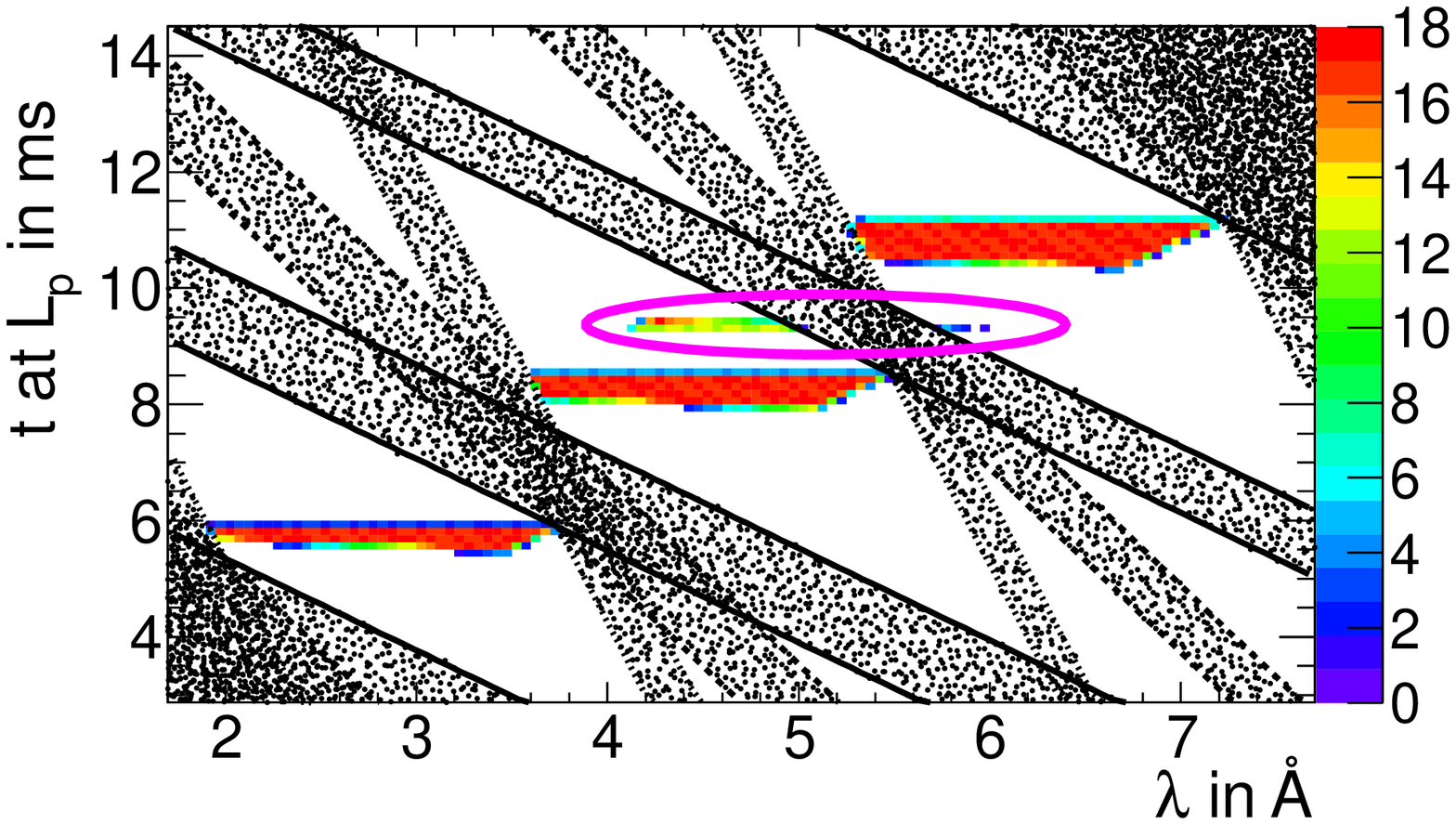}}
	\hfill
        \caption{Remaining phase space after subsequent inclusion of frame overlap choppers at 12.125\,m, 18.25\,m and 30.5\,m. Areas that are excluded by the FOCs are shaded. While there is hardly any contaminant radiation left for the design resolution of $2.2\%$, there is a clear leakage of spurious neutrons highlighted by the magenta ellipse into the second and third subframe when reducing the distance between the two discs of the PSC to achieve a resolution of $1\%$.}
\label{Fig:PhaseSpaceWithOldFOC}
 \end{center}
\end{figure}

The inclusion of FOCs restricts parts of the phase space transmitted through the PSCs (Fig. \ref{Fig:PhaseSpaceWithOldFOC}). This leads to a reduced transmission for wavelengths being in the overlap region of the individual subframes. The level of such a flux reduction also depends on other instrument parameters and is discussed in the next section, while this discussion is more focused on whether the FOCs keep all the unwanted phase space away from the subframes. While it appears that for the loosest resolution of $\delta \lambda / \lambda = 2.2\% $ the transmitted parameter space is in accordance with expectations, at a higher resolution  of $1\%$, when the discs of the PSCs are closer together, there is a leakage of phase space into subframes 2 and 3, which spoils the desired resolution. Thus the previously chosen layout of FOCs does not work properly for all adjustable WFM settings. \\

\begin{figure}
\begin{center}
        \subfigure[Phase space after all choppers for $2.2\%$ resolution.]{\includegraphics[width=0.45\textwidth,keepaspectratio]{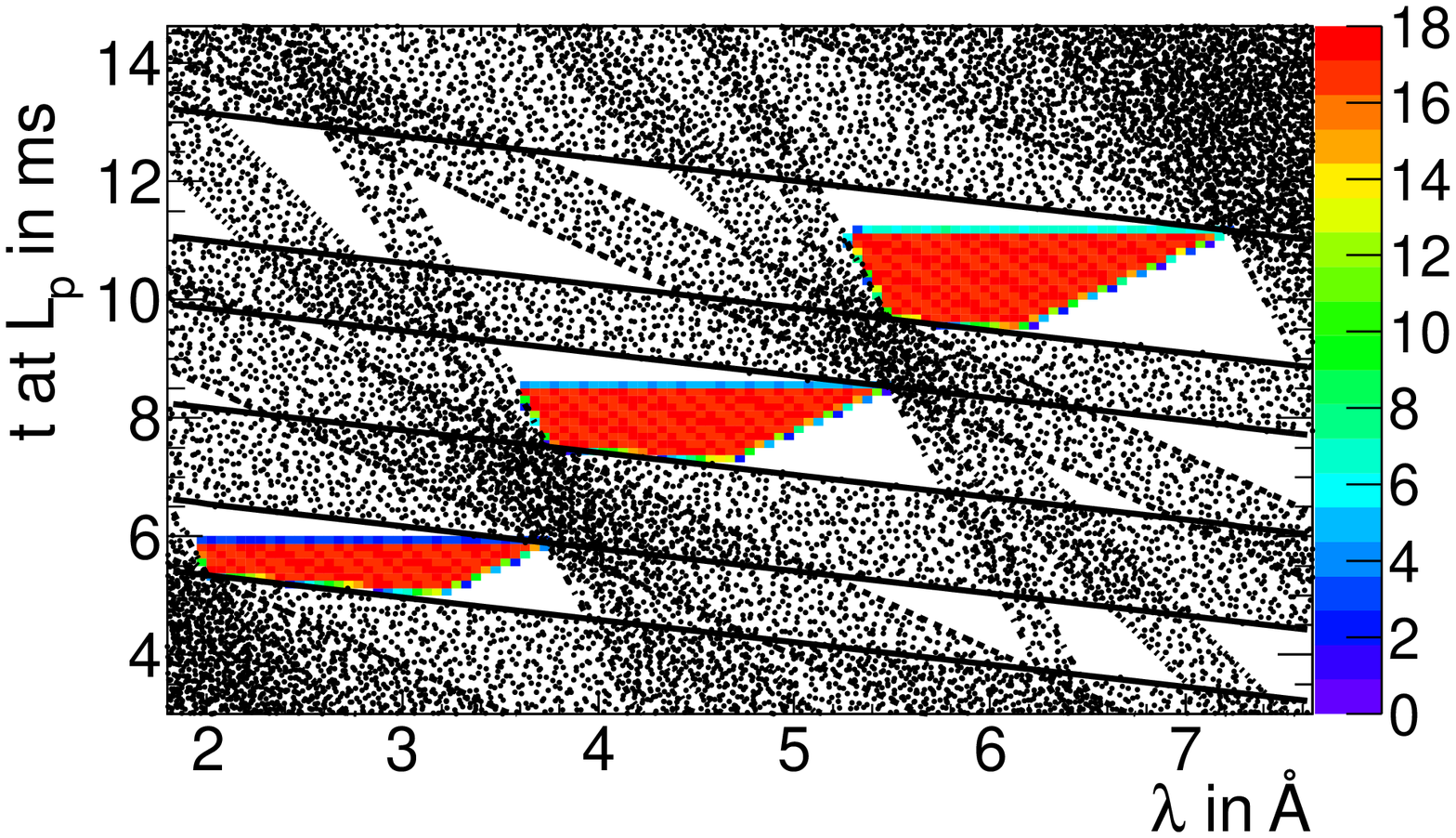}}
        \subfigure[Phase space after all choppers for $1\%$ resolution.]{\includegraphics[width=0.45\textwidth,keepaspectratio]{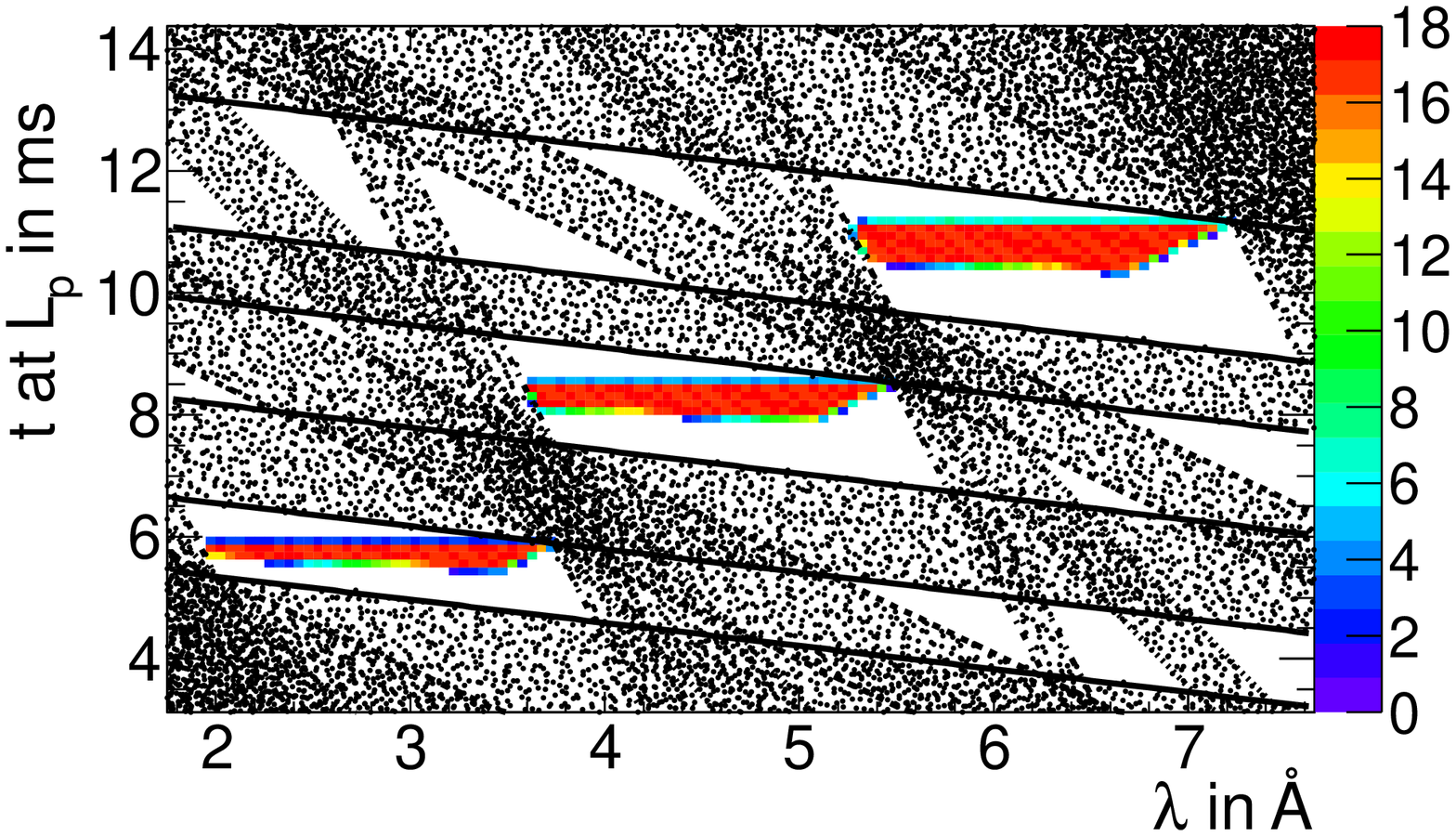}}
        \caption{The inclusion of a fourth FOC at 7.5\,m removes the contaminant radiation present in the WFM setup shown in Fig. \ref{Fig:PhaseSpaceWithOldFOC}. Now even for resolutions of $1\%$ (and higher) the transmitted phase space is free of spurious neutrons.}
\label{Fig:PhaseSpaceWithNewFOC} 
\end{center}
\end{figure}

\begin{figure}
\begin{center}
        \subfigure[Total ToF of neutrons at the detector position]{\includegraphics[width=0.485\textwidth,keepaspectratio]{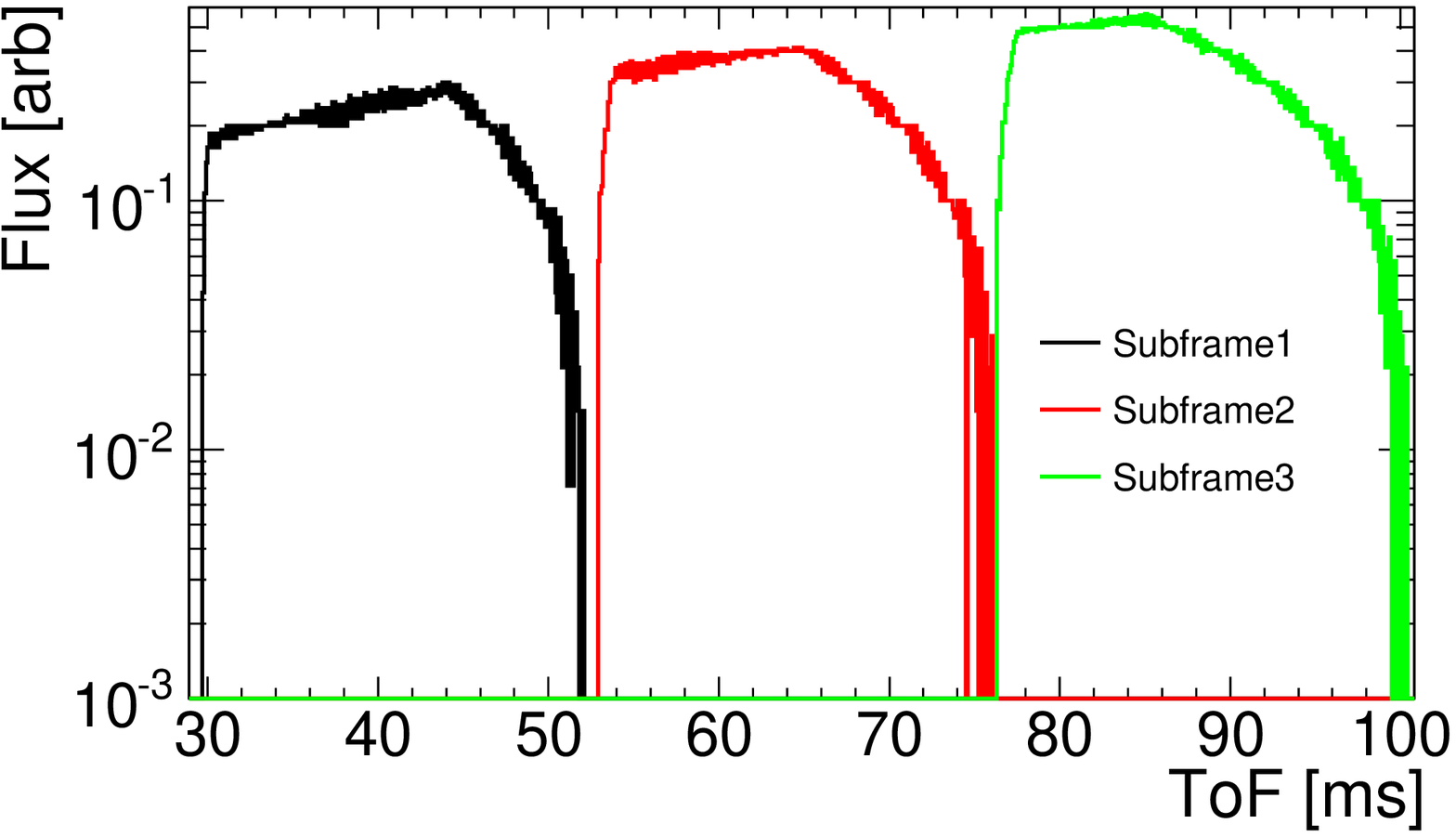}}
        \subfigure[Time resolution at the detector position]{\includegraphics[width=0.485\textwidth,keepaspectratio]{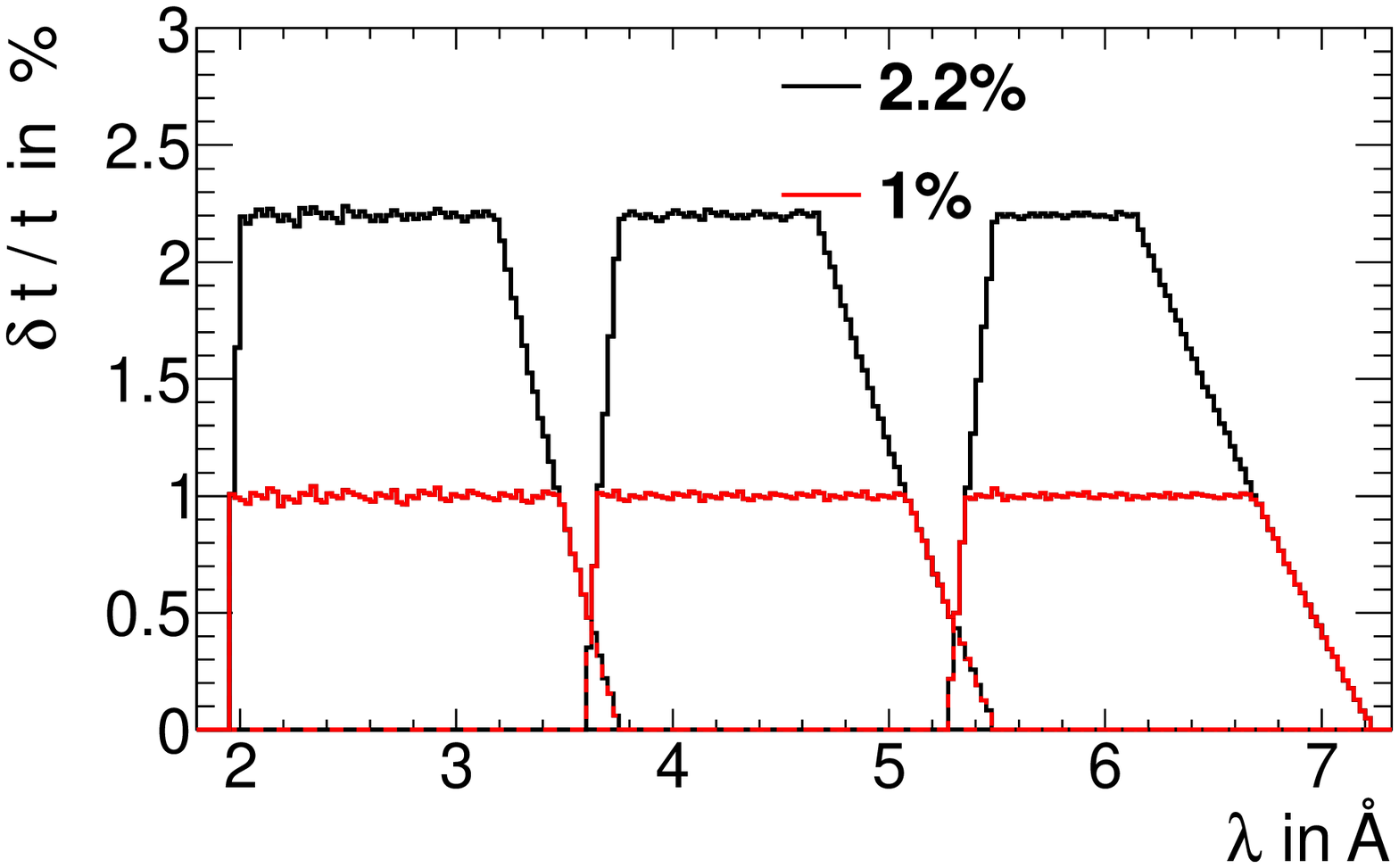}}
        \caption{Left: ToF plot of 3 subframes coming from a single main pulse, which are well separated in time. Right: The wavelength resolution at the detector expressed as $\delta t (\lambda) / t_{\mathrm{tot}} (\lambda)$, where $t_{\mathrm{tot}}$ is the ToF of neutrons between the centre of the PSC and the detector. The contributions of individual subframes are denoted by dashed lines, whereas the maximum resolution is depicted by the solid lines. Since the subframes are separated in time, it allows for an unambiguous reconstruction of the wavelength from ToF.}
\label{Fig:ToFWithNewFOC}
 \end{center}
\end{figure}

\begin{figure}
\begin{center}
	\includegraphics[width=0.8\textwidth]{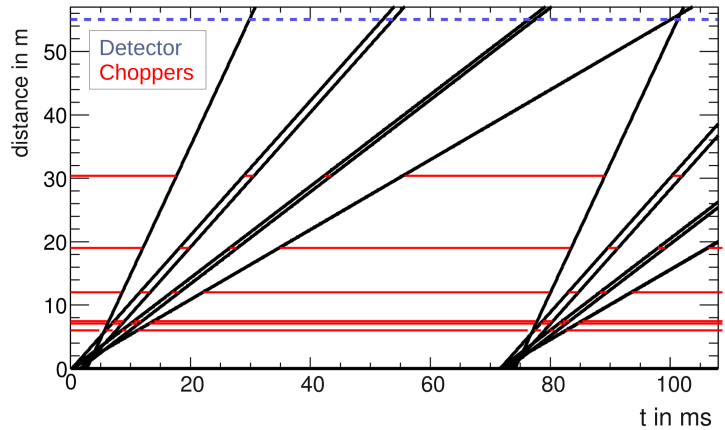}
	\caption{Time of flight diagram of the final chopper setup as worked out with the acceptance diagram method. The fastest and slowest of neutrons trajectories in the individual subframes are represented by black lines, while choppers and the detector are depicted by the red and blue lines, respectively. For completeness, the next main pulse is shown as well.}
\label{Fig:ToFComplete}
\end{center}
\end{figure}

The position of the contaminant phase space in the diagram suggests that an additional FOC located very close to the PSC, i.e. represented by lines with a very small slope, would be able to remove the frame overlap while at the same time not cut into the usable phase space. This is confirmed in Fig. \ref{Fig:PhaseSpaceWithNewFOC}, showing the addition of a FOC at 7.5\,m, while also the positions of other three FOCs were slightly changed (see Tab. \ref{Tab:LR} and Fig. \ref{Fig:ToFComplete}). Contaminant radiation is now removed even for high resolutions while saving as much as possible of the usable phase space. In Fig. \ref{Fig:ToFWithNewFOC}, analytical calculations of neutron propagation through this chopper setup show that all subframes are separated in time at the detector position, while the adjusted resolution is achieved for a greater part of the usable waveband. For wavelengths close to a neighbouring (sub)frame, the resolution and thus the transmission is reduced due to prevention of frame overlap. As the next step, the validity of this layout needs to be confirmed by neutronic Monte-Carlo (MC) simulations, described in the following section. \\

\section{Comparison with MC simulations}

The analytical study described in the last section makes use of idealised conditions. In a real instrument, the characteristics of the transmitted neutron beam will be influenced by additional parameters like guide geometry, beam divergence and pulse structure, and chopper rotation speed. Thus to confirm that the WFM chopper layout derived from analytical considerations is suitable for a real instrument, it needs to be tested by a neutron MC simulation, where all of these criteria are included. In this work, the VITESS software \cite{Bib:Vitess, Bib:VitessURL} package was used. The chopper setup was included in the simulations of the \textit{instrument I}, which will be published elsewhere. \\

\subsection{Simulations of the reflectometer chopper layout}

\begin{figure}
\begin{center}
	\includegraphics[width=0.8\textwidth]{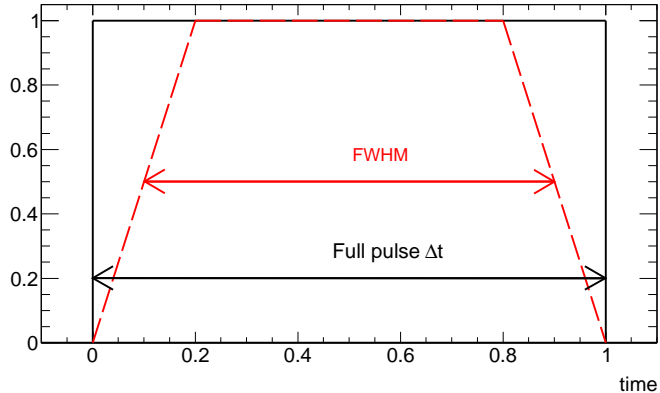}
	\caption{Illustration of the neutron pulse structure used in the analytical study and MC simulations. While in the analytical study the opening and closing time of choppers were assumed to be infinitely small and thus the pulse was a perfect rectangle with a width of $\Delta t (\lambda)$, the finite guide size and chopper rotation speed lead to a trapezoidal shape of the pulse. Its full width at half maximum (FWHM) is smaller than the pulse duration $\Delta t$, since in this work the points in time at which pulse starts and ends in the MC simulation were decided to exactly coincide with those from the phase space study.}
\label{Fig:Pulse}
\end{center}
\end{figure}

In order to include the choppers in the MC simulation, it is important to decide on their parameters like radius and rotation speed. The radius and rotation speed might be constrained by their position in the particular instrument and engineering feasibility. It is also important to decide how to deal with the finite time a chopper needs to fully open or close the beam. First, in order to be conservative and prevent frame overlap as far as possible, the time $t^O_{i,j}$ ($t^C_{i,j}$), at which the $i$th chopper opens (closes) the guide in the analytical calculation, is defined as the time at which the chopper \textrm{starts to open} (\textrm{fully closes}) the beam in the simulation, see Fig. \ref{Fig:Pulse}. This requirement guarantees that for each wavelength the neutron transmission starts and ends at the same time as in the phase space study. Hence the size of the windows has to be reduced to account for the time the choppers need to sweep through the guide. As a result, for a given nominal resolution simulations should yield a higher measured resolution at the cost of a reduced transmission due to a smaller FWHM of the pulse. A deviation from this strict requirement is considered in the next section.\\

\begin{figure}
\begin{center}
        \subfigure[Time resolution at the detector position for 2.2\%]{\includegraphics[width=0.485\textwidth,keepaspectratio]{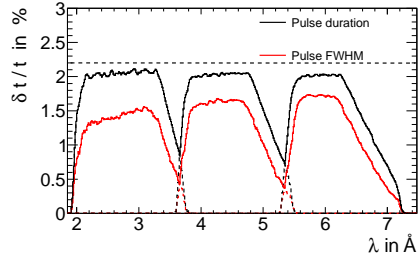}}
        \subfigure[Time distribution at the detector position for 2.2\%]{\includegraphics[width=0.485\textwidth,keepaspectratio]{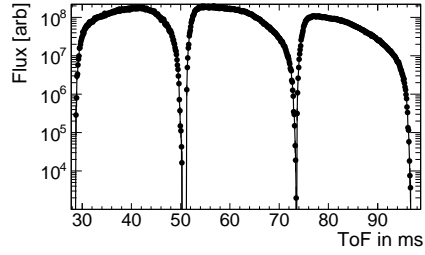}}
	\hfill
	\subfigure[Time resolution at the detector position for 1\%]{\includegraphics[width=0.485\textwidth,keepaspectratio]{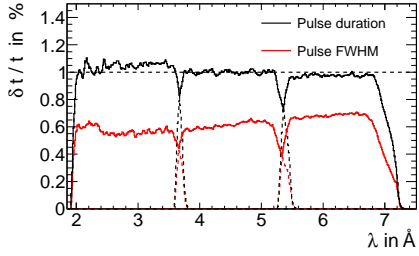}}
        \subfigure[Time distribution at the detector position for 1\%]{\includegraphics[width=0.485\textwidth,keepaspectratio]{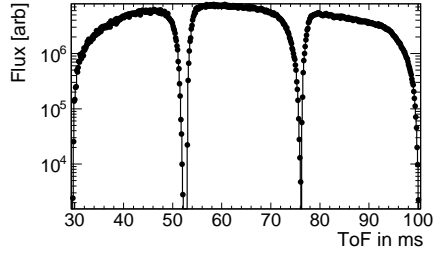}}
        \caption{(a) and (c): Measured time resolution at the detector position as a function of wavelength, which was calculated using both the total pulse duration $t_{\mathrm{max}} - t_{\mathrm{min}} (\lambda)$ and its FWHM (see also Fig. \ref{Fig:Pulse}). As expected, the total pulse duration agrees well with analytical results while for the FWHM calculation the trapezoidal shape of the pulses due to finite guide geometry and chopper rotation speed comes into play. (b) and (d): ToF distribution at the detector position, all subframes are clearly separated in time by the WFM setup.}
\label{Fig:PerfSimulations}
 \end{center}
\end{figure}

To prove that the WFM setup works in the MC simulation, it is important to show that both the desired resolution is reached and the subframes are well separated in time. Results of VITESS simulations shown in Fig. \ref{Fig:PerfSimulations} confirm that the subframes are well separated in time and the time gap between subframes coincides with analytical results. As far as the achieved time resolution is concerned, it can be observed that especially for short wavelengths it is higher than the nominal resolution, thus the neutron transmission is slightly worse in MC simulations compared with the transmission from analytical calculations. The wavelength spectrum exhibits dips as a result of frame overlap prevention, see Fig. \ref{Fig:Spectrum} and Fig. \ref{Fig:ToFWithNewFOC} and \ref{Fig:PerfSimulations} for comparison.

\subsection{Impact of technical constraints}

\begin{figure}
\begin{center}
	\includegraphics[width=0.8\textwidth]{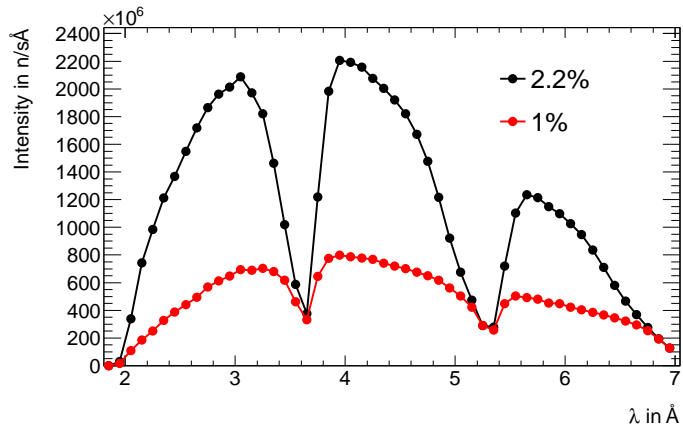}
	\caption{Neutron flux at the detector position for the \textit{instrument I} comprising a WFM chopper layout for $2.2\%$ and $1\%$ wavelength resolution. For wavelengths close to the subframe edges a reduction of flux due to frame overlap prevention can be observed.}
\label{Fig:Spectrum}
\end{center}
\end{figure}

\begin{figure}
\begin{center}
        \subfigure[Measured resolution at the detector position for nominal resolution of $2.2\%$]{\includegraphics[width=0.485\textwidth,keepaspectratio]{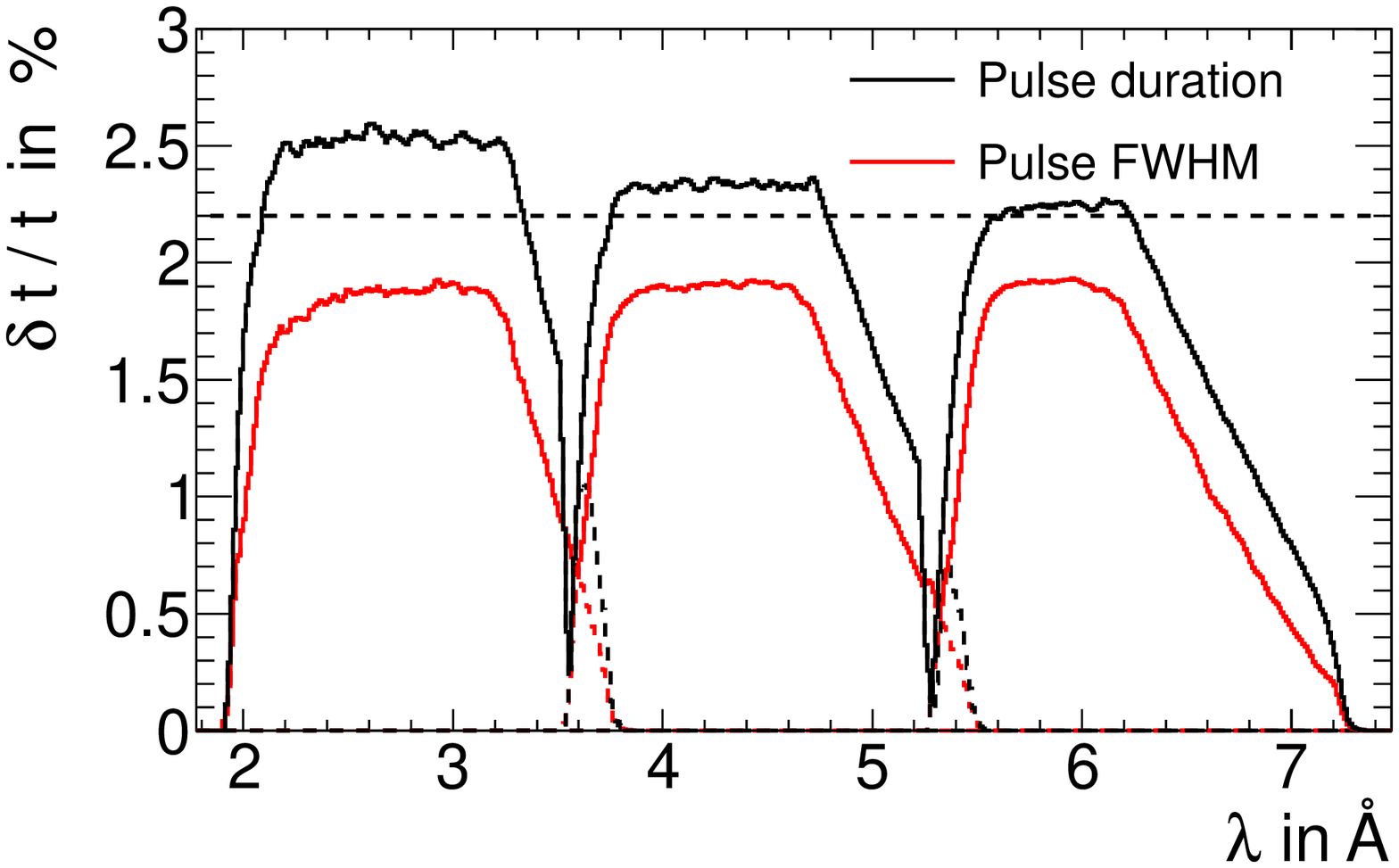}}
        \subfigure[Measured resolution at the detector position for nominal resolution of $1\%$]{\includegraphics[width=0.485\textwidth,keepaspectratio]{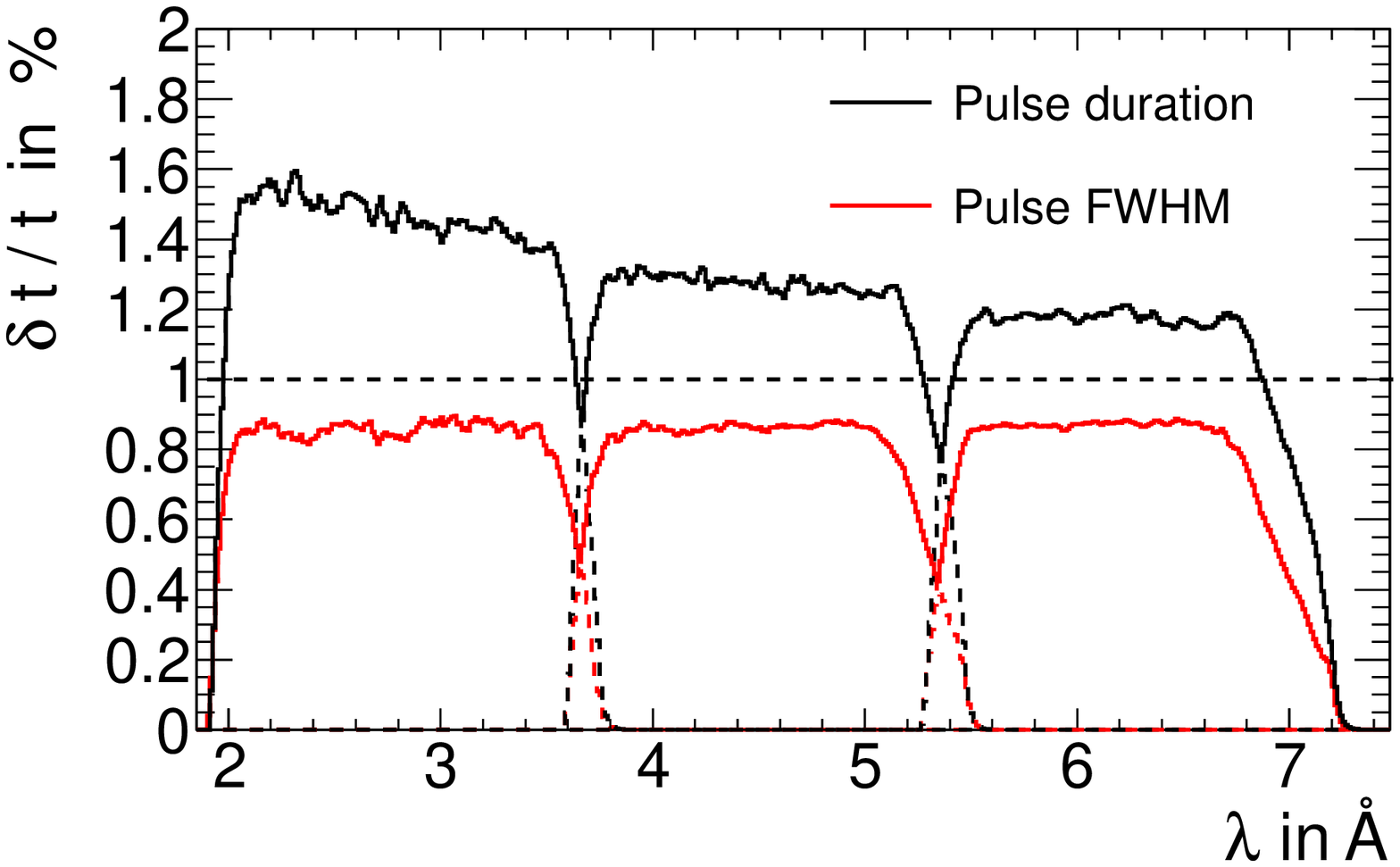}}
 \caption{Wavelength resolution measured at the detector position using the total width and FWHM of time pulses as a function of wavelength. The effect of reduced and wavelength dependent FWHM due to finite instrument geometry and chopper speed (see Fig. \ref{Fig:PerfSimulations}) is corrected by modifying the windows of the PSC. See text for further details.}
\label{Fig:PulseWidthsModified}
 \end{center}
\end{figure}

\begin{figure}
\begin{center}
	\includegraphics[width=0.8\textwidth]{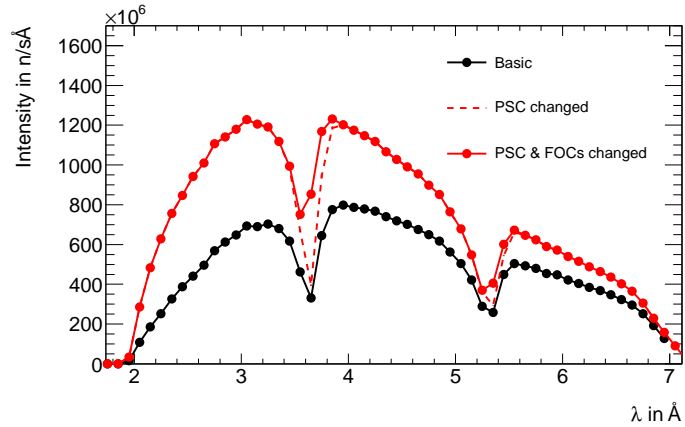}
	\caption{Neutron flux at the detector position for the \textit{instrument I} comprising a WFM chopper layout for $1\%$ wavelength resolution. The basic configuration of choppers, depicted by the black line, was modified to maximize the flux output in the regions where subframes overlap in wavelength. An improved performance was reached when modifying the windows of the PSC as well as those of FOCs.}
\label{Fig:SpectrumComparison}
\end{center}
\end{figure}

In the last section it was shown that the WFM setup as developed with the help of acceptance diagrams proved to work in the MC simulation of the \textit{instrument I}. Compared to analytical calculations, geometrical constraints of the instrument have an impact on the neutron transmission and lead to time pulses, which deviate from the idealised rectangular shape (see Fig. \ref{Fig:Pulse}). This has an effect on the achieved wavelength resolution (Fig. \ref{Fig:PerfSimulations}) and overall neutron flux (Fig. \ref{Fig:Spectrum}). As far as the resolution is concerned, in order to achieve the desired value either the distance between the discs of the PSC needs to be increased or the windows of the PSC should be modified. The latter can be done by withdrawing the reduction of the window widths that accounted for finite guide dimensions, i.e. dropping the strict requirement concerning chopper opening and closing times by assuming that the beam is infinitely thin. This leads to an increase of the total pulse width, but at the same time the FWHM of the pulse, which is the factor determining the wavelength resolution at the detector, better corresponds to the desired value, see Fig. \ref{Fig:PulseWidthsModified}. Such a choice of window parameters for the PSC can be recommended as a solution to the pulse shape problem coming from finite instrument dimensions. Flux losses in the regions around subframe edges, which come from FOCs cutting into the beam to avoid frame overlap, can be reduced by optimizing the sizes and offsets of chopper windows such that the time gap between subframes is minimised and the opening and closing time is reduced (see Fig. \ref{Fig:SpectrumComparison}). \\

\begin{figure}
\begin{center}
        \subfigure[Measured resolution at the detector position for nominal resolution of $1\%$]{\includegraphics[width=0.485\textwidth,keepaspectratio]{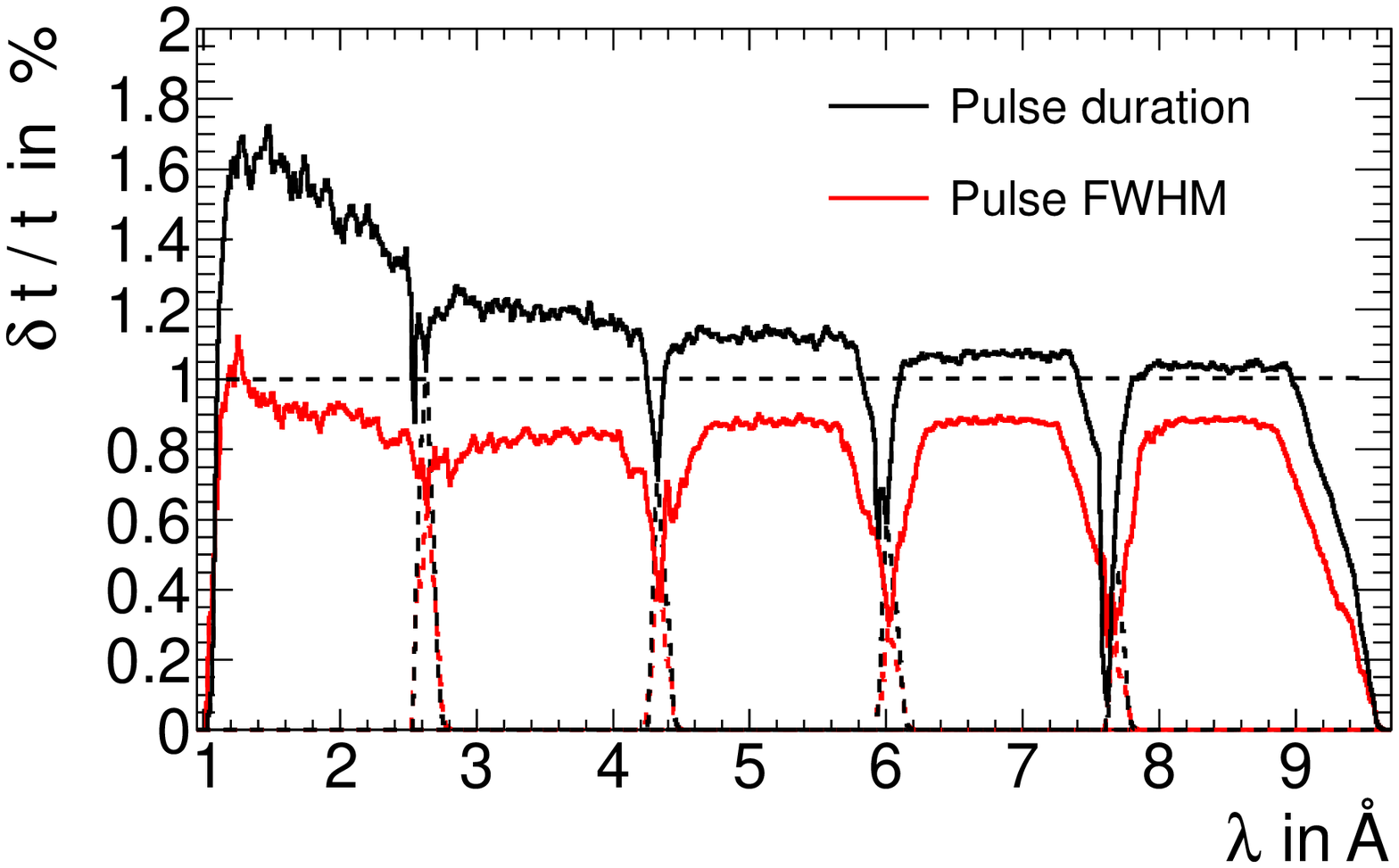}}
        \subfigure[Measured ToF at the detector position for nominal resolution of $1\%$]{\includegraphics[width=0.485\textwidth,keepaspectratio]{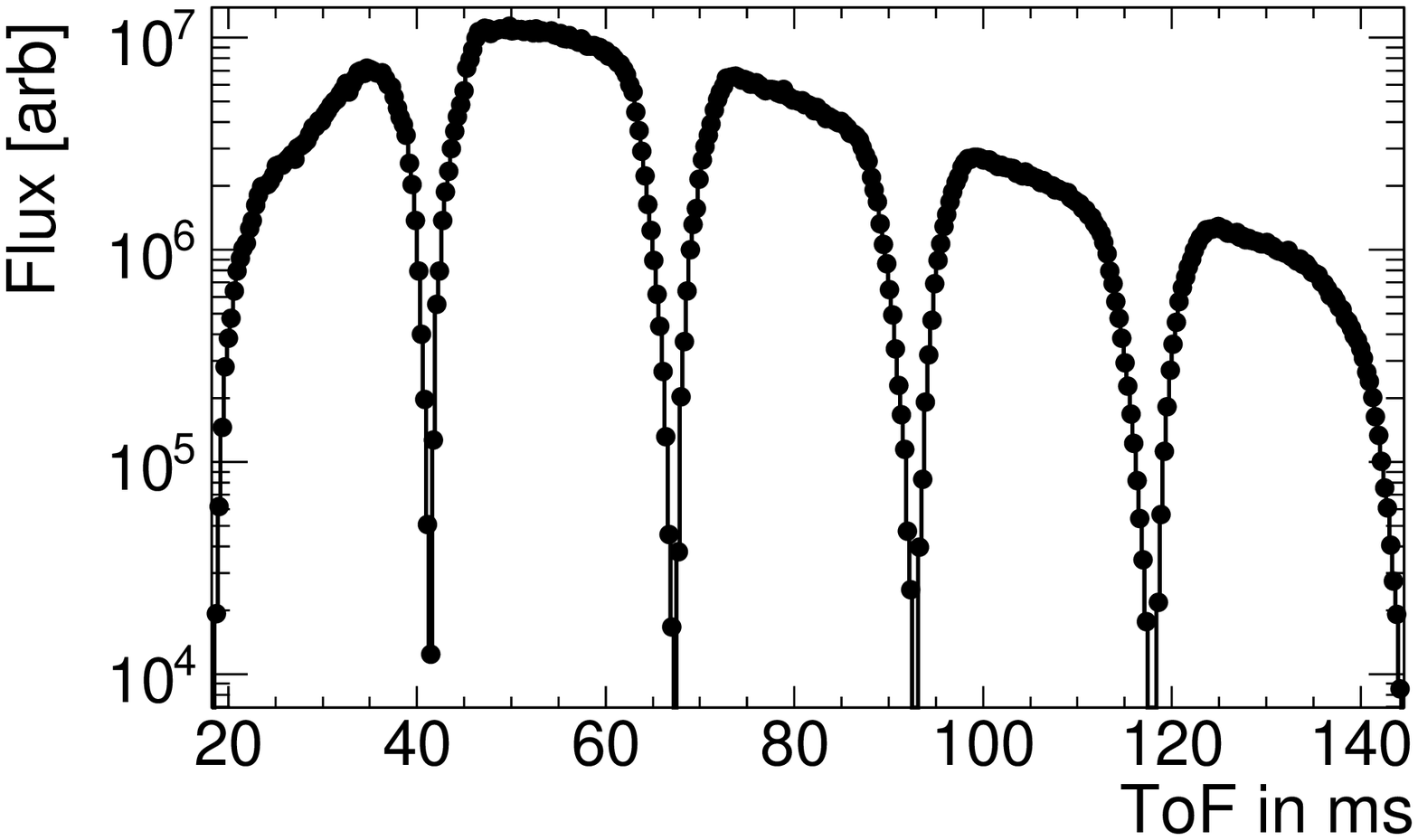}}
 \caption{Measured resolution and ToF distribution for the \textit{instrument II} having a $9\times9 \, \mathrm{cm}^2$ guide cross section for the most length and utilizing wavelengths between 1 and around 10 \, \AA. The chopper layout designed with acceptance diagrams allows to reach the adjusted resolution by splitting the waveband into five subframes that do not overlap in time.}
\label{Fig:Performance9cmInstrument}
 \end{center}
\end{figure}

\begin{figure}
\begin{center}
	\includegraphics[width=0.8\textwidth]{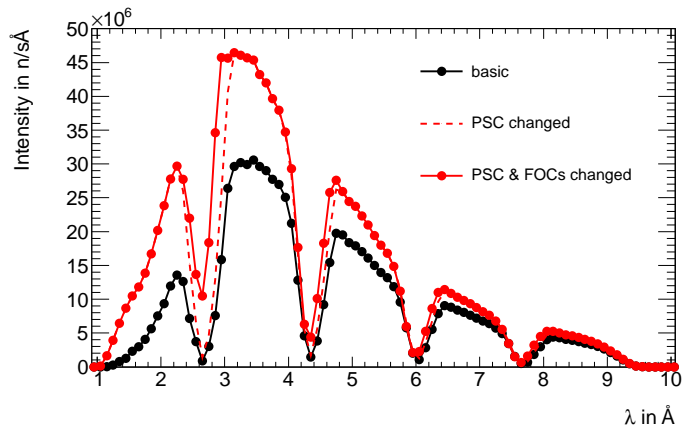}
	\caption{Neutron flux at the detector position for the \textit{instrument II} comprising a WFM chopper layout for $1\%$ wavelength resolution. The basic configuration of choppers, depicted by the black line, was modified to maximize the flux output in the regions where subframes overlap in wavelength. An improved performance was reached when modifying the windows of the PSC as well as those of FOCs. See text for further details.}
\label{Fig:SpectrumComparison9cm}
\end{center}
\end{figure}

It should be mentioned that the \textit{instrument I} does not have the most difficult conditions in terms of the complexity of the WFM system, both in terms of the used wavelength band and instrument geometry, in particular taking into account the small height of the neutron guide of 2\,cm. To prove that the concept still works in more challenging conditions as well, it was applied to a comparable instrument (\textit{instrument II}) requiring a constant resolution for wavelengths between $1 \, \mathrm{\AA}$ and about $10 \, \mathrm{\AA}$ and having a guide cross section of $9 \times 9 \, \mathrm{cm}^2$ for the most of the length of the instrument. The chopper layout worked out with acceptance diagrams was very similar to the one for \textit{instrument I}, again comprising six choppers and in particular with the first FOC being placed very close to the PSC, which is again located at 6\,m. While the PSC and the first FOC deal with a focused beam of a $2 \times 2 \, \mathrm{cm}^2$ cross section \footnote{If high-resolution measurements are desired, the instrument concept should be such that at the position of the PSC the beam is narrow at least in one dimension. Since at the future ESS there are tight space constraints for choppers placed at around 6\,m, a large beam cross section would render pulse shaping for high-resolution mode impossible.}, the full guide cross section of $9 \times 9 \, \mathrm{cm}^2$ is seen at the positions of the remaining three FOCs. MC simulations show that also in this case the chopper system delivers the desired resolution for the entire waveband, which is split into five subframes being all separated in time as required (Fig. \ref{Fig:Performance9cmInstrument}). The flux losses due to frame overlap avoidance increase, since the larger guide dimensions and smaller chopper speed due to the increased transmitted waveband require longer opening and closing chopper times than for the \textit{instrument I}. This situation can be improved by minimising the time gap between subframes (see Fig. \ref{Fig:SpectrumComparison9cm}). For this, acceptance diagrams once more prove to help by pointing out the right chopper parameters for a modification. Compared to the \textit{instrument I}, there is more flux lost in the overlap regions, however the total flux reduction only amounts to about $20\%$, if compared to a layout in which FOCs would be excluded. In general, the spectrum transmitted by a WFM system and its optimisation will be particular to each instrument, whereas at the same time a chopper layout suggested by the acceptance diagram approach can be expected to be already close to an optimum solution.

\section{Conclusion}

The WFM concept is a sophisticated chopper setup that enables to expand the usable wavelength range, in particular in combination with a constant wavelength resolution setup at long pulse neutron sources. Due to its complexity, the design of such a system is challenging and there are several criteria that need to be accounted for. As was shown in this work, acceptance diagrams can be a powerful tool to design and optimise WFM systems, because they help getting a thorough understanding of the interplay between individual choppers and are at the same time much faster to process than neutron simulations, thus problems like contaminant neutrons at higher resolutions would be more difficult to recognise and solve in MC simulations. Acceptance diagrams allow one to optimise the number and positions of the WFM choppers such that the beam characteristics obtained in MC simulations match the instrument requirements in terms of subframe separation and achieved resolution. The presented WFM concept works for different instruments independent of their particular geometrical constraints, thus the acceptance diagram method can be of significant help when designing or upgrading instruments, in particular in view of the future ESS facility. \\

\section*{Acknowledgements}

We thank M. Trapp, M. Strobl and R. Steitz for their fruitful discussions.\\
This work was funded by the German BMBF under ``Mitwirkung der Zentren der Helmholtz Gemeinschaft und der Technischen Universit\"at M\"unchen an der Design-Update Phase der ESS, F\"orderkennzeichen 05E10CB1.''

\clearpage{}

\clearpage{}

\begin{table}[htbp]
\begin{center}
  \begin{tabular}{|l|c|c|}
    \hline
    \textbf{ Parameter }& \multicolumn{2}{c|}{\textbf{ Parameter value }}   \\
    \hline
      & \textit{Instrument I} & \textit{Instrument II} \\	
    \hline	
    \hline	
     ESS pulse length $t_0$ & \multicolumn{2}{c|}{ $2.86 \, \mathrm{ms}$ }\\
    \hline
     ESS source frequency & \multicolumn{2}{c|} {$14 \, \mathrm{Hz}$} \\
    \hline
     Total instrument length $L_{\mathrm{tot}}$ &  $55 \, \mathrm{m}$ & $60 \, \mathrm{m}$ \\
    \hline
     Wavelength band  & 2--7.2 $\mathrm{\AA}$ & 1--9.6 $\mathrm{\AA}$  \\
    \hline
     Distance between the PSCs and detector $L_0$ & $49 \, \mathrm{m}$ & $54 \, \mathrm{m}$ \\	
     \hline
     Position of the first PSC & \multicolumn{2}{c|} {$6 \, \mathrm{m}$} \\	
    \hline
     Position of the second PSC at $2.2\%$ ($1\%$) resolution & $7.08 \, \mathrm{m}$ ($6.49 \, \mathrm{m}$) & --- ($6.54 \, \mathrm{m}$) \\	
    \hline
     Rotation frequency of the PSC & \multicolumn{2}{c|} {$70 \, \mathrm{Hz}$ } \\
    \hline	
     Final position and rotation frequency of the 1st FOC & $7.5 \, \mathrm{m}$, $70 \, \mathrm{Hz}$ & $7.4 \, \mathrm{m}$, $70 \, \mathrm{Hz}$\\
     \hline
     Final position and rotation frequency of the 2nd FOC & $12 \, \mathrm{m}$, $56 \, \mathrm{Hz}$ & $11.7 \, \mathrm{m}$, $42 \, \mathrm{Hz}$\\
     \hline
     Final position and rotation frequency of the 3rd FOC & $19 \, \mathrm{m}$, $28 \, \mathrm{Hz}$ & $18 \, \mathrm{m}$, $28 \, \mathrm{Hz}$\\
     \hline
     Final position and rotation frequency of the 4th FOC & $30.4 \, \mathrm{m}$, $14 \, \mathrm{Hz}$  & $28 \, \mathrm{m}$, $14 \, \mathrm{Hz}$\\
     \hline 	
     Guide height & $2 \, \mathrm{cm}$ & $2 - 9 \, \mathrm{cm}$ \\
     \hline
     Guide width & $10 - 26 \, \mathrm{cm}$ & $2 - 9 \, \mathrm{cm}$ \\
     \hline
  \end{tabular}
  \caption{Basic preliminary instrument parameters used in the design of the potential future ESS liquids reflectometer (\textit{instrument I}) and for the crosscheck instrument (\textit{instrument II}).}\label{Tab:LR}
\end{center}
\end{table}

\clearpage{}

\end{document}